\documentclass[10pt]{article}
\usepackage{amssymb}
\usepackage{amsmath}
\usepackage{amsthm}
\usepackage{latexsym}
\usepackage[dvips]{epsfig}
\usepackage{mathrsfs}
\usepackage[bookmarks,colorlinks=false]{hyperref}

\theoremstyle{plain}
\newtheorem{proposition}{Proposition}
\newtheorem{remark}{Remark}
\newtheorem{lemma}{Lemma}
\newtheorem{theorem}{Theorem}

\newtheorem{definition}{Definition}

\newcommand{\updn}[3]{#1^{#2}_{\phantom{#2}#3}}

\begin{document}

\title{\textbf{A local non-negative initial data scalar characterisation of the Kerr solution}}

\author{{\Large Alfonso Garc\'{\i}a-Parrado G\'omez-Lobo} \thanks{E-mail address:
{\tt alfonso@math.uminho.pt}} \\
Centro de Matem\'atica, Universidade do Minho,\\
4710-057 Braga, Portugal.\footnote{Current address: F\'{\i}sica Te\'orica, Universidad del Pa\'{\i}s Vasco,
Apartado 644, 48080 Bilbao, Spain}}
\maketitle

\begin{abstract}
For any vacuum initial data set, we define a local, non-negative scalar quantity which vanishes at every point 
of the data hypersurface if and only if the data are {\em Kerr initial} data. Our scalar quantity only depends on the quantities used 
to construct the vacuum initial data set which are the Riemannian metric defined 
on the initial data hypersurface and a symmetric tensor which plays the role of the second fundamental form of the embedded initial data hypersurface.
The dependency is {\em algorithmic} in the sense that given the initial data one can compute the scalar quantity by algebraic and differential 
manipulations, being thus suitable for an implementation in a numerical code. The scalar could also be useful in studies of the non-linear stability of the 
Kerr solution because it serves to measure the deviation of a vacuum initial data set from the Kerr initial data in a local and algorithmic way.
\end{abstract}

PACS: 04.20.Jb, 95.30.Sf, 04.20.-q\\

MSC: 83C15, 83C57, 83C05 

\section{Introduction}
The Kerr metric \cite{KERR-METRIC} is one of the most important exact solutions in general relativity and it has been extensively studied since its discovery 
both from the mathematical and physical point of view. On the physical side the solution represents a rotating black hole with the physical properties that
black holes existing in nature are expected to have and therefore the Kerr solution is regarded as a good approximation to rotating black holes in an 
astrophysical scenario \cite{TEUKOLSKYKERR}.

On the mathematical side we know that this solution is the subject of powerful existence and uniqueness theorems 
(see \cite{CHRUSCIEL-COSTA} for an up-to-date 
revision of them) and it is also the subject of important open questions in mathematical general relativity. One of them is the expectation that 
the Kerr solution is the asymptotic final state in the evolution of an isolated system which has undergone gravitational collapse turning 
it into a black hole. Other open issue is how far the existence and uniqueness results mentioned above can be refined when assumptions such as the 
analiticity of the space-time are dropped (the most recent results in this direction can be found in 
\cite{IONESCU-KL-ALEXAKIS,ALEXAKISSMALLROTATING,IONESCUKLAINERMANREVIEW}). 
There is also the 
old problem of the existence of an interior solution which can be matched to the Kerr solution through a hypersurface playing the role of 
the surface of a body. Finally a fourth open question is the {\em non-linear stability} of the {\em domain of outer communication} of the Kerr solution. 
The non-linear stability of solutions of the Einstein's field equations is a subject which has received recently wide attention 
and a number of different techniques to address 
this important problem have been employed (a selection of them 
can be found in \cite{FRI86A,FRI86B,FRI91,CHRKLA93,KLAINNICO99,LINROD05,KLAINERMANAXIALKERR}).

A recent approach towards the non-linear stability problem relies on the construction of non-negative quantities out of a vacuum initial data set which
are zero if and only if the data development is a subset of a globally hyperbolic region of the Kerr geometry 
\cite{KERR-INVARIANT-TJ,KERR-INVARIANT-PRL, KERR-INVARIANT-PRS, NON-KERRNESS-COMPACT}. The idea is that these quantities could serve to define 
when a vacuum initial data set is close to a {\em Kerr initial data set}. Such a notion is necessary in order to formulate the problem of the 
non-linear stability of the Kerr solution. With a similar motivation it is also possible to introduce non-negative quantities characterizing 
the Kerr solution defined in the space-time rather than on an initial data set \cite{GARSENKERRNESS}. In the present work we follow 
this motivation and show how to 
construct a non-negative scalar out of the quantities used to define a vacuum initial data set which vanishes if and only if the data set is a 
{\em Kerr initial data set}. We follow the convention laid by previous authors and call the scalar 
we construct the {\em non-Kerrness} 
even though this 
does not necessarily imply that our scalar is related in some way to other previous definitions. 

In our approach the non-Kerrness can be defined on a 3-dimensional Riemannian manifold playing the role of a vacuum initial data set 
for the Einstein's field equations.
What this means is that there exists a symmetric tensor field on the manifold which together with the Riemannian metric fulfills 
the so-called vacuum constraint conditions. These conditions are satisfied if and only if the Riemannian manifold can be isometrically embedded in a 4-dimensional spacetime which is 
a vacuum solution of the Einstein equations. The image of the Riemannian manifold under the embedding corresponds 
to the initial data hypersurface 
of the vacuum equations and the symmetric tensor is just the second fundamental form of the embedded manifold. 
The non-Kerrness defined in this work only depends on the Riemannian metric, the symmetric tensor used to define 
the vacuum initial data set and their covariant
derivatives (with respect to the Riemannian metric Levi-Civita connection), all of them evaluated at each point 
of the Riemannian manifold. It is a non-negative scalar computed in an algorithmic fashion and
we prove that this scalar is zero at every point of the Riemannian manifold if and only if it is isometrically 
embeddable into the Kerr solution. The algorithmic nature just pointed out implies that 
if the scalar is used in a numerical simulation then its computation is not 
very demanding once the data on a numerical slice are known.

This paper is structured as follows: the mathematical preliminaries and notation used are summarised in section 
\ref{preliminaries}. In section \ref{sec:ferrando-saez} we review the local characterisation of the Kerr 
solution developed by Ferrando \& S\'aez \cite{FERSAEZKERR}. This is a characterisation of the Kerr solution written 
in terms of {\em concomitants} of the Riemann tensor and it plays an essential role in our work. Ferrando \& S\'aez
characterisation is reviewed in Theorem \ref{theorem:kerr-characterization}. 
In section \ref{sec:decomposition} we perform the {\em orthogonal splitting} of the conditions given in Theorem 
\ref{theorem:kerr-characterization} and obtain a number of results needed to define our {\em local non-Kerrness}. 
The orthogonal splitting of Ferrando \& S\'aez conditions yield necessary conditions which ensure that the development of 
a vacuum initial initial data set is a subset of the Kerr space-time. In principle one needs additional conditions to 
obtain a characterisation of a Kerr initial data set. To obtain these we follow a procedure already used in \cite{GARVALSCH} which 
consists in finding the conditions that guarantee the existence of a Killing vector in the data development. Once this is done we use the 
Killing vector to propagate the necessary conditions obtained in the previous step to the data development, thus proving that they hold 
in an open set of the space-time which cointains the initial data hypersurface. The necessary and sufficient conditions required for 
the existence of a Killing
vector in the data development are reviewed in Theorem \ref{thm:kid} and in Lemma \ref{lemma:omega} we explain how to use them in our proof.
With all this information we construct in section \ref{sec:positive-scalar} the non-negative scalar quantity used to characterise a Kerr initial data set. 
The scalar quantity is defined in Theorem \ref{theorem:kerr-initial-data} where we prove that the scalar has the required properties.
Finally we discuss some applications in section \ref{sec:conclusions}. The computations carried out in this paper have been performed with the {\em xAct} 
tensor computer algebra suite \cite{XACT}.

\section{Preliminaries} \label{preliminaries}
Let $(\mathcal{M},g_{\mu\nu})$ denote a smooth orientable
spacetime. Unless otherwise stated all the structures defined on $\mathcal{M}$ are assumed to be smooth.
The following conventions will be used: plain Greek letters
$\alpha,\beta,\gamma,\dots$ denote abstract indices in four dimensions
and, occasionally, free index notation will be used in which case 
we will denote the corresponding tensors with boldface letters (and an arrow overhead if the tensor is a 
vector field on the manifold).
The signature of the metric tensor
$g_{\mu\nu}$ will be taken to be $(-,+,+,+)$,
while $R_{\mu\nu\alpha\rho}$,
$R_{\mu\nu}=R^{\alpha}_{\phantom{\alpha}\mu\alpha\nu}$ and
$C_{\alpha\beta\mu\nu}$ denote, respectively the Riemann, Ricci and
Weyl tensors of $g_{\mu\nu}$. The tensor $\eta_{\alpha\beta\sigma\nu}$
is the volume element which is used to define the Hodge dual of any
antisymmetric tensor. This operation is denoted by the standard way of
appending a star $*$ to the tensor symbol. 
The operator $\pounds_{\vec{\boldsymbol u}}$
symbolizes the Lie derivative with respect to the vector field
$\vec{\boldsymbol u}$. 

\subsection{The notion of Kerr initial data}
\label{subsec:cauchy-problem}
We review next the standard formulation of the Cauchy problem for
the vacuum Einstein equations. In this formulation one considers a
3-dimensional connected Riemannian manifold $(\Sigma,h_{ij})$ ---we
use small plain Latin letters $i,j,k,\dots$ for the abstract indices
of tensors on this manifold--- and an isometric embedding
$\phi:\Sigma\longrightarrow\mathcal{M}$. The map $\phi$ is an
isometric embedding if $\phi^*g_{\mu\nu}=h_{ij}$ where as usual
$\phi^*$ denotes the pullback of tensor fields from $\mathcal{M}$ to
$\Sigma$.  The metric $h_{ij}$ defines a unique affine connection
$D_i$ (Levi-Civita connection) by means of the standard condition
\begin{equation}
D_jh_{ik}=0.
\end{equation}
The Riemann tensor of $D_i$ is $r_{ijkl}$ and from it we define its
Ricci tensor by $r_{ij}\equiv r^{l}_{\phantom{l}ilj}$ and its scalar
curvature $r\equiv r^i_{\phantom{i}i}$ ---in $\Sigma$ indices are
raised and lowered with $h_{ij}$ and its inverse $h^{ij}$.
\begin{theorem} 
Let $(\Sigma,h_{ij})$ be a Riemannian manifold and suppose that
there exists a symmetric tensor field $K_{ij}$ on it which satisfies
the conditions (vacuum constraints)
\begin{subequations}
\begin{eqnarray}
&& r+ K^2-K^{ij}K_{ij}=0, \label{Hamiltonian}\\
&& D^jK_{ij}-D_iK=0, \label{Momentum}
\end{eqnarray}
\end{subequations}  
where $K\equiv K^{i}_{\phantom{i}i}$. Provided that $h_{ij}$ and $K_{ij}$ are 
smooth there exists an isometric embedding $\phi$ of
$\Sigma$ into a globally hyperbolic, vacuum solution
$(\mathcal{M},g_{\mu\nu})$ of the Einstein field equations. The set
$(\Sigma,h_{ij}, K_{ij})$ is then called a vacuum initial data
set and the spacetime $(\mathcal{M},g_{\mu\nu})$ is the data
development. Furthermore the spacelike hypersurface
$\phi(\Sigma)$ is a Cauchy hypersurface in $\mathcal{M}$.
\label{theorem:vacuum-data}
\end{theorem}
As stated before we assume the smoothness of $h_{ij}$, $K_{ij}$ in the previous theorem but it is in fact true under more general 
differentiability assumptions (see Theorem 8.9 of \cite{CHOQUETBRUHATBOOK}).
Since $\phi(\Sigma)$ is a Cauchy hypersurface of $\mathcal{M}$ we shall often use the standard notation
$D(\Sigma)$ for $\mathcal{M}$ (the identification $\Sigma\leftrightarrow\phi(\Sigma)$ is then
implicitly understood).

The main object of this paper is the characterisation of {\em Kerr initial data}. We 
give next the formal definition of this concept.

\begin{definition}
 A vacuum initial data set $(\Sigma, h_{ij}, K_{ij})$ is called Kerr initial data or 
 a Kerr initial data set if there exists an isometric embedding 
 $\phi:\Sigma\rightarrow\mathcal{M}$ where $\mathcal{M}$ is an open subset of the Kerr spacetime.
\label{def:kerr-initial-data}
\end{definition}

By the {\em Kerr space-time} we understand the maximal extension of the Kerr solution \cite{CARTER-KERR-EXTENSION}.
We do not impose any restriction on the mass or angular momentum parameters of the Kerr solution. 
In principle the open subset of the previous definition can be in any region of the Kerr space-time although in the applications
one is interested in globally hyperbolic regions of the Kerr space-time.

Under the conditions of the Theorem \ref{theorem:vacuum-data} and the Definition 
\ref{def:kerr-initial-data}, we may construct a foliation of $\mathcal{M}$ 
with $n^\mu$ as the timelike unit vector which is orthogonal to the leaves.
We shall denote by $\{\Sigma_t\}_{t\in I\subset\mathbb{R}}$ the family of leaves of
this foliation and we choose the foliation in such a way that $\phi(\Sigma)=\Sigma_0$
(we assume that $0\in I$).
The leaf $\Sigma_0$ is called the {\em initial data hypersurface}. The interest of introducing 
a foliation is that the unit vector $n^\mu$ enables us to perform the {\em orthogonal} splitting
of any tensorial quantity defined on $\mathcal{M}$ and relate the parts resulting from this splitting
to tensorial quantities defined on $\Sigma$ by means of the embedding $\phi$. This is a well-known 
procedure which we review next for the sake of clarity in our exposition.

\subsubsection{The orthogonal splitting}\label{section:osplitting}
Let $n^\mu$ be a unit timelike vector, $n^\mu n_\mu=-1$ defined on
$\mathcal{M}$. Then any tensor can be
decomposed with respect to $n^\mu$ and the way to achieve it is the
essence of the orthogonal splitting (also known as 3+1 formalism)
which is described in many places of the literature ---see e.g.
\cite{ELLIS-COSMOLOGY,NATARIOCOSTAANALOGY} and references therein). We review next the parts of this formalism needed
in this work. The spatial metric is defined by $h_{\mu\nu}\equiv
g_{\mu\nu}+n_\mu n_\nu$ and it has the algebraic properties
$h^\mu_{\phantom{\mu}\mu}=3$,
$h_\mu^{\phantom{\mu}\sigma}h_{\sigma\nu}=h_{\mu\nu}$. We shall call a
covariant tensor $T_{\alpha_1\dots \alpha_m}$ {\em spatial} with
respect to $h_{\mu\nu}$ if it is invariant under
$h^\mu_{\phantom{\mu}\nu}$ i.e. if
\begin{equation}
h^{\alpha_1}_{\phantom{\alpha_1}\beta_1}\cdots h^{\alpha_m}_{\phantom{\alpha_m} \beta_m}T_{\alpha_1\cdots \alpha_m}=T_{\beta_1\cdots\beta_m}, 
\end{equation}
with the obvious generalization for any mixed tensor. This property
is equivalent to the inner contraction of $n^\mu$ with $T_{\alpha_1\dots
\alpha_m}$ (taken on any index) vanishing.  The orthogonal splitting
of a tensor expression consists in writing it as a sum of terms which
are tensor products of the unit normal and spatial tensors of lesser
degree or the same degree in which case the unit normal is absent (see eq. (\ref{eq:tensor-general-ot})).
In order to find the orthogonal splitting of expressions containing
covariant derivatives we need to introduce the {\em spatial
  derivative} $D_\mu$ which is an operator whose action on any tensor
field $\updn{T}{\alpha_1\dots\alpha_p}{\beta_1\dots\beta_q}$,
$p,q\in\mathbb{N}$ is given by
\begin{equation}
D_\mu \updn{T}{\alpha_1\dots\alpha_p}{\beta_1\dots\beta_q}\equiv
h^{\alpha_1}_{\phantom{\alpha_1}\rho_1}\dots h^{\alpha_p}_{\phantom{\alpha_p} \rho_p}h^{\sigma_1}
_{\phantom{\sigma_1}\beta_1}\dots
h^{\sigma_q}_{\phantom{\sigma_q}\beta_q}h^{\lambda}_{\phantom{\lambda}\mu}
\nabla_\lambda
T^{\rho_1\dots\rho_p}_{\phantom{\rho_1\dots\rho_p}\sigma_1\dots\sigma_q}.
\label{spatialcd}
\end{equation}
From (\ref{spatialcd}) is clear that $D_\mu \updn{T}{\alpha_1\dots\alpha_p}{\beta_1\dots\beta_q}$ is spatial.

The results just described hold for an arbitrary unit timelike vector
$n^\mu$ but in our framework we only need to consider integrable
timelike vectors which are characterized by the condition
$n_{[\mu}\nabla_{\nu}n_{\sigma]}=0$ (Frobenius condition). 
In this case there exists a
foliation of $\mathcal{M}$ such that the vector field $n^\mu$ is
orthogonal to the leaves of the foliation.  We shall work with foliations related to an initial data set 
$(\Sigma, h_{ij}, K_{ij})$ by means of an imbedding $\phi:\Sigma\rightarrow\mathcal{M}$ as
explained after Definition \ref{def:kerr-initial-data}.

The tensor $h_{\mu\nu}$ plays
the role of the {\em first fundamental form} for any of the leaves
while the symmetric tensor $K_{\mu\nu}$ defined by
\begin{equation}
K_{\mu\nu}\equiv-\frac{1}{2}\mathcal{L}_{\vec{n}}h_{\mu\nu},
\label{lie-h}
\end{equation}
can be identified with the {\em second fundamental form} for any of the leaves. 
Combining previous definition with the Frobenius condition we easily derive
\begin{equation}
\nabla_\mu n_\nu=-K_{\mu\nu}-n_\mu A_\nu,
\label{gradnormaltoextrinsic}
\end{equation}
where $A^\mu\equiv n^\rho\nabla_\rho n^\mu$ is the acceleration of $n^\mu$. 

By using the quantities just introduced one is in principle able to work out the orthogonal splitting  
of any tensorial expression. If $T_{\mu_1\dots\mu_p}$, $p\in\mathbb{N}$ is an arbitrary covariant 
tensor field on $\mathcal{M}$ then we can write its orthogonal splitting in the form 
\begin{equation}
{\boldsymbol T}=\sum_{J,P} {\boldsymbol T}^{(J)}_{(P)} {\boldsymbol n}_{J},
\label{eq:tensor-general-ot}
\end{equation}
where ${\boldsymbol n}_J$ represents a product of $J$-copies of the 1-form $n_\mu$ with appropriate abstract indices 
and ${\boldsymbol T}^{(J)}_{(P)}$ is a spatial covariant tensor with respect
to $n_\mu$. The index $P$ labels all possible spatial tensors appearing in the splitting.
If no factors $n_\mu$ are present then we set $J=0$. 

Once the orthogonal splitting of a tensor in $\mathcal{M}$ has been achieved
then the terms of the splitting are directly related to tensors in $\Sigma$ by means of the embedding
$\phi$. For example one has $\phi^*h_{\mu\nu}=h_{ij}$. 
Other key relations are
\begin{equation}
\phi^*K_{\mu\nu}=K_{ij},\ 
\phi^*(D_\mu T_{\beta_1\dots\beta_q})=
D_i(\phi^*T_{\beta_1\dots\beta_q}).
\end{equation}
In general for any tensor ${\boldsymbol T}$ defined on $\mathcal{M}$ we have the property
\begin{equation}
{\boldsymbol T}|_{\phi(\Sigma)}=0\Longleftrightarrow {\boldsymbol T}^{(J)}_{(P)}=0 \Longleftrightarrow
\phi^*({\boldsymbol T}^{(J)}_{(P)})=0  \;,\forall J\;,\quad
\label{eq:spatialp}
\end{equation}
where the pullback $\phi^*({\boldsymbol T}^{(J)}_{(P)})$ of any spatial covariant tensor ${\boldsymbol T}^{(J)}_{(P)}$ is computed by just 
replacing the Greek indices by Latin ones.

We supply below the explicit orthogonal splittings for some particular space-time tensors 
which will be needed repeatedly in the text
(see for example \cite{SE-DYNAMICALAWS} and references therein 
for further details about these formulae).
\begin{itemize}

 \item Orthogonal splitting of the volume element
\begin{equation}
\eta_{\alpha\beta\gamma\delta}=-n_\alpha\varepsilon_{\beta\gamma\delta}+n_\beta
\varepsilon_{\alpha\gamma\delta}
-n_\gamma\varepsilon_{\alpha\beta\delta}+n_\delta\varepsilon_{\alpha\beta\gamma}.
\label{decompose_eta}
\end{equation}
Here $\varepsilon_{\alpha\beta\gamma}\equiv n^\mu\eta_{\mu\alpha\beta\gamma}$ is the {\em spatial volume element} which is a fully antisymmetric spatial tensor.

\item Orthogonal splitting of a symmetric tensor. If ${\mathcal S}_{\mu\nu}$ is an 
arbitrary symmetric tensor then its orthogonal splitting takes the form 
\begin{equation}
{\mathcal S}_{\mu\nu}=\mathfrak{r}({\mathcal S}) n_\mu n_\nu+2\mathfrak{j}({\mathcal S})_{(\mu}n_{\nu)}+\mathfrak{t}({\mathcal S})_{\mu\nu}\;,
\end{equation}
where $\mathfrak{j}({\mathcal S})_\mu$ and $\mathfrak{t}({\mathcal S})_{\mu\nu}$ are spatial and 
$\mathfrak{t}({\mathcal S})_{(\mu\nu)}=\mathfrak{t}({\mathcal S})_{\mu\nu}$.

\item 
Orthogonal splitting of a {\em Weyl candidate}.
Any tensor,
$W_{\mu\nu\lambda\rho}$, with the same algebraic pro\-per\-ties as the Weyl tensor can be decomposed into its 
\emph{electric part}, $\mathcal{E}(W)_{\mu\nu}$, and \emph{magnetic part},
$\mathcal{B}(W)_{\mu\nu}$. More precisely, one has that
\begin{subequations}
\begin{eqnarray}
&&W_{\mu\nu\lambda\sigma}=2\left(l_{\mu[\lambda}\mathcal{E}(W)_{\sigma]\nu}-l_{\nu[\lambda}\mathcal{E}(W)_{\sigma]\mu}
-n_{[\lambda}\mathcal{B}(W)_{\sigma]\tau}\varepsilon^\tau_{\phantom{\tau}\mu\nu}
-n_{[\mu}\mathcal{B}(W)_{\nu]\tau}\varepsilon^\tau_{\phantom{\tau}\lambda\sigma} 
\right)\;,\nonumber\\
&&\label{weyl-1generalized} \\
&&W^*_{\mu\nu\lambda\sigma}=2\left(l_{\mu[\lambda}\mathcal{B}(W)_{\sigma]\nu}-l_{\nu[\lambda} \mathcal{B}(W)_{\sigma]\mu}
+n_{[\lambda}\mathcal{E}(W)_{\sigma]\tau}\varepsilon^\tau_{\phantom{\tau}\mu\nu}
+n_{[\mu}\mathcal{E}(W)_{\nu]\tau}\varepsilon^\tau_{\phantom{\tau}\lambda\sigma}\right)\;,\nonumber\\
&&\label{weyl-2generalized}
\end{eqnarray}
\end{subequations}
where
\begin{equation}
\mathcal{E}(W)_{\tau\sigma}\equiv W_{\tau\nu\sigma\lambda}n^\nu n^\lambda, \quad 
\mathcal{B}(W)_{\tau\sigma}\equiv W^*_{\tau\nu\sigma\lambda}n^\nu n^\lambda, \quad 
[\mathcal{E}(W)_{\tau\sigma}]=[\mathcal{B}(W)_{\tau\sigma}]=L^{-2}
\label{eq:e-m-generic}
\end{equation}
and $l_{\mu\nu}\equiv h_{\mu\nu}+n_{\mu}n_{\nu}$, while
$\varepsilon_{\tau\lambda\sigma}\equiv\varepsilon_{\nu\tau\lambda\sigma}n^{\nu}$.
The tensors $\mathcal{E}(W)_{\mu\nu}$ and $\mathcal{B}(W)_{\mu\nu}$
are symmetric, traceless, and spatial: $n^\mu
\mathcal{E}(W)_{\mu\nu}=n^\mu \mathcal{B}(W)_{\mu\nu}=0$.

\item Orthogonal splitting of a Riemann-like tensor. If $\mathcal{R}_{\mu\nu\rho\lambda}$ is 
a tensor with the same algebraic symmetries as the Riemann tensor then we define
\begin{equation}
\ ^{*}\mathcal{R}_{\alpha\beta\gamma\delta}\equiv\frac{1}{2}\eta_{\alpha\beta\rho\sigma}\mathcal{R}^{\rho\sigma}_{\phantom{\rho\sigma}\gamma\delta},\ 
\mathcal{R}^{*}_{\alpha\beta\gamma\delta}\equiv\frac{1}{2}\eta_{\rho\sigma\gamma\delta}\mathcal{R}_{\alpha\beta}^{\phantom{\alpha\beta}\rho\sigma},\ 
\ ^{*}\mathcal{R}^*_{\alpha\beta\gamma\delta}\equiv\frac{1}{2}\eta_{\alpha\beta}^{\phantom{\alpha\beta}\rho\sigma}\mathcal{R}^*_{\rho\sigma\gamma\delta}.
\end{equation}

Next we introduce the following spatial tensors (spatial parts of $\mathcal{R}_{\mu\nu\rho\lambda}$)
\begin{equation}
Y(\mathcal{R})_{\alpha\gamma}\equiv \mathcal{R}_{\alpha\beta\gamma\delta}n^\beta n^\delta,\ 
Z(\mathcal{R})_{\alpha\gamma}\equiv \ ^{*}{\mathcal R}_{\alpha\beta\gamma\delta}n^\beta n^\delta,\ 
X(\mathcal{R})_{\alpha\gamma}\equiv\ ^{*}{\mathcal R}^*_{\alpha\beta\gamma\delta}n^\beta n^\delta.
\label{eq:riemann-parts}
\end{equation}
The algebraic symmetries of the Riemann tensor entail the properties
\begin{equation}
X(\mathcal{R})_{(\alpha\beta)}=X(\mathcal{R})_{\alpha\beta},\ Y(\mathcal{R})_{(\alpha\beta)}=Y(\mathcal{R})_{\alpha\beta},
\ Z(\mathcal{R})^\alpha_{\phantom{\alpha}\alpha}=0.
\label{eq:riemann-parts-prop}
\end{equation}
If just the monoterm symmetries of the Riemann tensor are assumed for $\mathcal{R}_{\mu\nu\alpha\beta}$ and the Bianchi
identity is not then the tensor $Z(\mathcal{R})_{\alpha\gamma}$ is not traceless.
The spatial parts of $\mathcal{R}_{\mu\nu\rho\lambda}$ contain all the information in $\mathcal{R}_{\mu\nu\rho\lambda}$ as is easily checked
by a simple count of their total number of independent components. They also enable
us to find the orthogonal splitting of $\mathcal{R}_{\mu\nu\alpha\beta}$ which reads
\begin{eqnarray}
&&R_{{\alpha\beta\gamma\delta}}=2n_\gamma n_{[\alpha}Y(\mathcal{R})_{{\beta]\delta}}+2h_{{\alpha[\delta}}X(\mathcal{R})_{{\gamma]\beta}}
+2n_\delta n_{[\beta} Y(\mathcal{R})_{{\alpha]\gamma}}+2n_{[\delta} Z(\mathcal{R})_{{\phantom{\rho}\gamma]}}^{{\rho}}\varepsilon_{{\alpha\beta\rho}}
+\nonumber\\
&&2n_{[\beta} Z(\mathcal{R})_{{\phantom{\rho}\alpha]}}^{{\rho}}\varepsilon_{{\gamma\delta\rho}}+h_{{\beta\delta}}\left(h_{{\alpha\gamma}} 
X(\mathcal{R})_{{\phantom{\rho}\rho}}^{{\rho}}-X(\mathcal{R})_{{\alpha\gamma}}\right)+h_{{\beta\gamma}}
\left(X(\mathcal{R})_{{\alpha\delta}}-h_{{\alpha\delta}} X(\mathcal{R})_{{\phantom{\rho}\rho}}^{{\rho}}\right).\nonumber\\
&&\label{eq:riemann-split}
\end{eqnarray}

\end{itemize}

\subsection{Killing initial data sets}
\label{subsec:kid}
A Killing initial data set (KID) associated to a vacuum initial data 
$(\Sigma, h_{ij}, K_{ij})$ is a pair
$(Y,Y_i)$ consisting of a scalar $Y$ and a vector $Y_i$ defined on
$\Sigma$ satisfying the following system of partial differential
equations on $\Sigma$ ---\emph{the KID equations}---
\begin{subequations}
\begin{eqnarray}
&& D_{(i}Y_{j)}-Y K_{ij}=0, \label{kid-1}\\ 
&&\hspace{-1cm} D_iD_jY-\mathcal{L}_{Y^l} K_{ij} =Y(r_{ij}+
KK_{ij}-2K_{il}K^l_{\phantom{l}j})\label{kid-2}.
\end{eqnarray}
\end{subequations} 
For us the most important result about a KID is the following theorem 
\cite{CHRUSCIEL-BEIG-KID,COLL77,MONCRIEF-KID}
\begin{theorem}
The necessary and sufficient condition for there to exist a Killing vector $\xi^\mu$ in the data development
of a vacuum initial data set $(\Sigma, h_{ij}, K_{ij})$ is that a pair $(Y, Y_j)$
fulfills eqs. (\ref{kid-1})-(\ref{kid-2}). The orthogonal splitting of $\xi^\mu$ 
with respect to any unit timelike normal $n^\mu$ related to the data set $(\Sigma, h_{ij}, K_{ij})$
as described in subsection \ref{subsec:cauchy-problem} reads 
\begin{equation}
 \xi_\mu= \tilde{Y} n_\mu+ \tilde{Y}_\mu\;,\quad \tilde{Y}\equiv -(n_\nu\xi^\nu)\;,\quad 
 \tilde{Y}_\mu \equiv (h_\mu{}^\nu\xi_\nu)\;,\quad 
 \phi^*(\tilde{Y})=Y\;,\quad \phi^*\tilde{Y}_\mu=Y_j.
\end{equation} 
\label{thm:kid}
\end{theorem}

From any arbitrary pair of covariant tensor fields $Y$, $Y_j$ on $\Sigma$, not necessarily fulfilling (\ref{kid-1})-(\ref{kid-2}), 
we define the following tensors for later use
\begin{equation}
\mathfrak{B}_{ij}\equiv D_{(i}Y_{j)}-Y K_{ij}\;,\quad
\mathfrak{C}_{ij}\equiv D_i D_j Y-\pounds_{Y^l} K_{ij} - Y(r_{ij}+ KK_{ij}-2K_{il}K^l_{\phantom{l}j}).
\label{eq:kid-scalars}
\end{equation}

\section{The invariant characterisation of Ferrando \& S\'aez}
\label{sec:ferrando-saez}
In this section we summarise the invariant characterisation of
the Kerr metric given by Ferrando \& S\'aez \cite{FERSAEZKERR}. To this
end we start by introducing some necessary nomenclature.

Let
$C_{\mu\nu\lambda\rho}$ denote the Weyl tensor of the
spacetime. Recall that
${}^*C_{\mu\nu\lambda\rho}=C^*_{\mu\nu\lambda\rho}$, a property that indeed holds for any 
Weyl candidate. Let
\[
G_{\mu\nu\lambda\rho}= g_{\mu\lambda} g_{\nu\rho} - g_{\mu\rho}g_{\nu\lambda}.
\] 

We shall make use of the following scalar invariants of the Weyl tensor:
\begin{eqnarray}
&& A \equiv \frac{1}{8} C^{\mu\nu\lambda\rho} C_{\mu\nu\lambda\rho},\\
&& B \equiv \frac{1}{8} C^{\mu\nu\lambda\rho} C^*_{\mu\nu\lambda\rho},\\
&& D \equiv \frac{1}{16} \updn{C}{\mu\nu}{\lambda\rho} \updn{C}{\sigma\pi}{\mu\nu} \updn{C}{\lambda\rho}{\sigma\pi},\\
&& E \equiv \frac{1}{16} \updn{C}{\mu\nu}{\lambda\rho} \updn{C}{\sigma\pi}{\mu\nu} \updn{C}{*\lambda\rho}{\sigma\pi}. 
\end{eqnarray}
Furthermore, let 
\begin{eqnarray}
&& \alpha = -\frac{AD+BE}{A^2+B^2}\;, \label{eq:define-alpha}\\
&& \beta=\frac{AE-BD}{A^2+B^2}.\label{eq:define-beta}
\end{eqnarray}
Next we define the following Riemann-like tensors
\begin{eqnarray}
&& Q_{\mu\nu\lambda\rho} \equiv \beta C_{\mu\nu\lambda\rho} + \alpha C^*_{\mu\nu\lambda\rho}\;, \quad \\
&& \Pi_{\mu\nu\lambda\rho} \equiv \alpha C_{\mu\nu\lambda\rho} -\beta C^*_{\mu\nu\lambda\rho} -(\alpha^2+\beta^2)G_{\mu\nu\lambda\rho}\;,
\end{eqnarray}
and the 1-forms
\begin{eqnarray}
&& R_\mu \equiv \frac{1}{3(A^2+B^2)}\left( A\nabla_\mu A + B\nabla_\mu B \right)\;, \\
&& \Theta_\mu \equiv  \frac{1}{3(A^2+B^2)}\left( B\nabla_\mu A - A\nabla_\mu B \right).
\end{eqnarray}
One needs further scalars
\begin{eqnarray}
&& K \equiv \frac{2 R_\mu \Theta^\mu}{R_\mu R^\mu -\Theta_\mu \Theta^\mu}\;,\\
&& T \equiv \frac{\beta}{\alpha}=\frac{BD-AE}{AD+BE}\;, \label{eq:define-T}\\
&& \lambda \equiv \frac{K(KT+1)}{K^2-3KT-2}\;,\label{eq:define-lambda} \\
&& \sigma \equiv \frac{2\alpha}{3\lambda^2-1}- \frac{R_{\mu}R^{\mu}-\Theta_{\mu}\Theta^{\mu}}{4(\lambda^2-1)}.
\end{eqnarray}
Using the above objects one defines
\begin{equation}
\Xi_{\mu\nu} \equiv \frac{1}{(1-\lambda^2)\sqrt{A^2+B^2}}
\big(\Pi_{\mu\lambda\nu\rho}(R^\lambda R^\rho -\Theta^\lambda \Theta^\rho) 
+ Q_{\mu\lambda\nu\rho} (R^\lambda \Theta^\rho + \Theta^\lambda R^\rho)\big).
\label{eq:xidefinition}
\end{equation}

\bigskip
The crucial result for our purposes is the following one (\cite{FERSAEZKERR}, Theorem 2).

\begin{theorem}
A solution $(\mathcal{M}, g_{\mu\nu})$ of the vacuum Einstein field equations is locally isometric to the Kerr spacetime if 
and only if
the following conditions hold in an open set of $\mathcal{M}$
\begin{eqnarray}
&& A^2+ B^2 \neq 0\;, \quad Q_{\mu\nu\lambda\rho} \nabla^\mu B \nabla^\lambda B\neq 0\;,\quad \sigma >0\;,\label{eq:signconditions} \\
&& \frac{1}{2}\updn{C}{\sigma\tau}{\mu\nu}C_{\sigma\tau\lambda\rho} + \alpha C_{\mu\nu\lambda\rho}+
\beta C^*_{\mu\nu\lambda\rho}- \frac{1}{3}\left( A G_{\mu\nu\lambda\rho}-
B\eta_{\mu\nu\lambda\rho}\right)=0, \label{eq:typeDcondition}\\
&& Q_{\mu\nu\lambda\rho} \nabla^\mu A \nabla^\lambda A + Q_{\mu\nu\lambda\rho} \nabla^\mu B \nabla^\lambda B=0, \\
&& (1-3\lambda^2)\beta + \lambda (3-\lambda^2)\alpha =0\;,
\end{eqnarray}
and there exists a vector field $\xi^\mu$ fulfilling the properties 
\begin{equation}
\Xi {}_{\mu}{}_{\nu}=\left(\frac{\alpha}{1 -3 \lambda^{2}}\right)^{\frac{2}{3}} 
\xi {}_{\mu} \xi {}_{\nu}\;,
\quad \nabla_\mu\xi_\nu+\nabla_\nu\xi_\mu=0.
\label{eq:stationarykilling}
\end{equation}
\label{theorem:kerr-characterization}
\end{theorem}

\begin{remark}\em
 The characterisation presented in Theorem \ref{theorem:kerr-characterization} is written in terms of 
 {\em concomitants} of the Weyl tensor. A concomitant of a tensor is a scalar or tensorial quantity which is constructed 
 exclusively from the tensor itself and its covariant derivatives with respect to the Levi-Civita connection.
\end{remark}

\begin{remark}\em
Condition (\ref{eq:stationarykilling}) was formulated 
in a slightly different way in \cite{FERSAEZKERR}
\begin{equation}
2 \Xi_{\mu\nu}u^\mu u^\nu+\Xi^{\mu}_{\phantom{\mu}\mu} > 0\;, 
\label{eq:ferr-saez-39}
\end{equation}
where $u^\mu$ is an arbitrary unit timelike vector. 
To check that indeed Theorem 2 of \cite{FERSAEZKERR} 
can be formulated with (\ref{eq:ferr-saez-39}) 
replaced by (\ref{eq:stationarykilling}) we first replace 
the value of $\Xi_{\mu\nu}$ shown in (\ref{eq:stationarykilling}) 
into the r.h.s of (\ref{eq:ferr-saez-39}) getting 
\begin{equation}
\left(\frac{\alpha}{1 -3 \lambda^{2}}\right)^{\frac{2}{3}} 
\left(2(u^\mu\xi_\mu)^2+\xi^\mu\xi_\mu\right).
\end{equation}
This scalar quantity is always positive if $u^\mu$ is unit timelike (this is trivial
if $\xi^\mu$ is space-like or null and follows from the reverse Cauchy-Schwarz inequality
when $\xi^\mu$ is timelike). 
Hence we conclude that any spacetime 
in which (\ref{eq:signconditions})-(\ref{eq:stationarykilling}) hold is 
locally isometric to the Kerr solution. 
Consistency with the Killing condition on $\xi^\mu$ follows 
by either Proposition 4 of \cite{FERSAEZKERR}
or by doing the explicit computation in a local coordinate chart of the Kerr metric.
Reciprocally, take the Kerr solution in the local coordinates given 
by (see \cite{EXACTSOLUTIONS}, eq. (21.16)) 
\begin{equation}
ds^2= \frac{(x^2 + y^2)}{X(x)}dx^2 + \frac{(x^2 + y^2)}{Y(y)}dy^2
 + \frac{1}{x^2 + y^2}\big(X(x)(dt -  y^2 dz)^2 - Y(y)(dt + x^2 dz)^2\big)\;,
\label{eq:kerr-metric}
\end{equation}
where 
\begin{equation}
 Y(y)\equiv\epsilon y^2 - 2 \mu y + \gamma\;,\quad 
 X(x) \equiv -\epsilon x^2 + \gamma\;,
\end{equation}
and $\epsilon$, $\gamma$, $\mu$ are positive constants. 
After doing the explicit computations in the coordinates of (\ref{eq:kerr-metric})
we can check that (\ref{eq:signconditions})-(\ref{eq:stationarykilling}) are all met. In this case the vector $\xi^\mu$ is given by
\begin{equation}
 \vec{\boldsymbol\xi}=-\frac{\sqrt{2}}{\mu^{1/3}}\frac{\partial}{\partial t}\;,
\end{equation}
which is obviously a Killing vector and so it fulfills the Killing condition.
According to the present analysis, 
the Killing condition on $\xi^\mu$ is in fact a consequence 
of the other conditions of Theorem \ref{theorem:kerr-characterization} in 
Ferrando \& S\'aez result, so it is somehow redundant. 
However, we chose to formulate the result including also this condition because it
is more adapted to our purposes.

\end{remark}

For the sake of completeness we also give the following result
also found in \cite{FERSAEZKERR}.

\begin{proposition}
\label{proposition:parameters}
If $g_{\mu\nu}$ is isometric to the Kerr metric, then:
\begin{itemize}
\item[(i)] the mass parameter, $m$, is given by
\begin{equation}
m\equiv \frac{|\alpha|}{\sigma \sqrt{\sigma} |3\lambda^2-1|}\;,
\label{eq:massparameter}
\end{equation}
\item[(ii)] the angular momentum parameter, $a$, is given by
\begin{equation}
a = \frac{1}{2\sigma (1+\lambda^2)}\left( \lambda R_\mu \Theta^\mu + \Theta_\mu \Theta^\mu + 4 \sigma \lambda^2   \right)^{1/2}.
\label{eq:angularmomentum}
\end{equation}
\end{itemize}
\end{proposition}

\section{Orthogonal splitting of Ferrando \& S\'aez characterisation}
\label{sec:decomposition}
The local characterisation of the Kerr solution presented in Theorem \ref{theorem:kerr-characterization} is 
written exclusively in terms of concomitants of the Weyl tensor and hence we can take each of these concomitants and 
compute their orthogonal splitting with respect to a unit vector orthogonal to the leaves of a foliation, in 
the manner explained in section \ref{section:osplitting}

We start by adapting (\ref{weyl-1generalized}) to the particular case of the Weyl tensor. In this case we
shall write
\begin{equation}
E_{\mu\nu} \equiv \mathcal{E}(C)_{\tau\sigma}, \quad B_{\mu\nu} \equiv \mathcal{B}(C)_{\tau\sigma}.
\label{eq:weyl-electric-magnetic}
\end{equation}
Note also that
\[
\mathcal{E}(W^*)_{\mu\nu}= B_{\mu\nu}, \quad \mathcal{B}(W^*)_{\mu\nu}=-E_{\mu\nu}.
\]

Crucial for the present purposes is to realize that given an initial data set $(\Sigma, h_{ij}, K_{ij})$ 
for the Einstein vacuum equations, the pull-back to $\Sigma$ 
of the spacetime tensors $E_{\mu\nu}$, $B_{\mu\nu}$ can be written entirely in terms of initial 
data quantities. 
Following the notation of \ref{section:osplitting}, if we set 
$E_{ij}\equiv \phi^*E_{\mu\nu}$ and $B_{ij}\equiv\phi^*B_{\mu\nu}$ then one has that
\begin{eqnarray}
&& E_{ij}= r_{ij} + K K_{ij} - K_{ik} \updn{K}{k}{j}\;, \label{eq:eb-to-initial-data0} \\
&& B_{ij}= \updn{\epsilon}{kl}{(i} D_{|k} K_{l|j)}.
\label{eq:eb-to-initial-data}
\end{eqnarray}
Therefore, any concomitant of the Weyl tensor whose orthogonal splitting results in an expression which 
only contains $E_{\mu\nu}$ and $B_{\mu\nu}$ (with no covariant derivatives thereof) can be written exclusively in terms of quantities defined 
from the data $(\Sigma, h_{ij}, K_{ij})$ when pull-backed to $\Sigma$. 
This will always be the case for a concomitant of the Weyl tensor not containing any covariant 
derivative of the Weyl tensor.

For later use we also need the evolution equations for $E_{\mu\nu}$ and $B_{\mu\nu}$ with respect to $\vec{\boldsymbol n}$ which 
result from the orthogonal splitting of the 
Bianchi identity of the Weyl tensor in vacuum. These are 
\begin{eqnarray}
&&\pounds_{\vec{\boldsymbol n}} E_{\mu\nu} = - 2A^{\beta} (B_{(\nu}{}^{\delta} \varepsilon_{\mu)\beta\delta}) 
+ 2 E_{\mu\nu} K^{\beta}{}_{\beta} - 2 E_{\nu}{}^{\beta} K_{\mu\beta} 
- 3 E_{\mu}{}^{\beta} K_{\nu\beta} + E^{\beta\delta} K_{\beta\delta} h_{\mu\nu} +
\varepsilon_{\nu\beta\delta} D^{\delta}B_{\mu}{}^{\beta},\nonumber\\
&&\label{eq:liedE}\\
&&\pounds_{\vec{\boldsymbol n}} B_{\mu\nu} = 2 A^{\beta} (E_{(\nu}{}^{\delta} \varepsilon_{\mu)\beta\delta}) 
+ 2 B_{\mu\nu} K^{\beta}{}_{\beta} - 2 B_{\nu}{}^{\beta} K_{\mu\beta} - 3 B_{\mu}{}^{\beta} K_{\nu\beta} + B^{\beta\gamma} K_{\beta\gamma} h_{\nu\mu} 
- \varepsilon_{\nu\beta\gamma} D^{\gamma}E_{\mu}{}^{\beta},\nonumber\\
&&\label{eq:liedB}
\end{eqnarray}

\bigskip
Next, using the orthogonal splitting of the Weyl tensor we write the scalars $A$, $B$, $D$, $E$ in terms of $E_{\mu\nu}$ and $B_{\mu\nu}$
\begin{eqnarray}
&& A=-B{}_{\mu}{}_{\nu} B{}^{\mu}{}^{\nu} + E{}_{\mu}{}_{\nu} E{}^{\mu}{}^{\nu}\;,\label{eq:splitting-A}\\
&& B=2 B{}^{\mu}{}^{\nu} E{}_{\mu}{}_{\nu}\;,\label{eq:splitting-B} \\
&& D= E{}_{\nu}{}_{\lambda}(3 B{}_{\mu}{}^{\lambda} B{}^{\mu}{}^{\nu}-E{}_{\mu}{}^{\lambda} E{}^{\mu}{}^{\nu})\;,\label{eq:splitting-D}\\
&& E=B{}^{\mu}{}^{\nu}(B{}_{\mu}{}^{\lambda} B{}_{\nu}{}_{\lambda} -3 E{}_{\mu}{}^{\lambda} E{}_{\nu}{}_{\lambda}).\label{eq:splitting-E}
\end{eqnarray}
Using (\ref{eq:liedE})-(\ref{eq:liedB}) on the first and second equations of the previous list we obtain
\begin{eqnarray}
&&\mathcal{A}(A)\equiv\pounds_{\vec{\boldsymbol n}} A = 6( B_{\alpha}{}^{\gamma} B^{\alpha\beta} -  E_{\alpha}{}^{\gamma} E^{\alpha\beta}) K_{\beta\gamma} + 
4 A K^{\gamma}{}_{\gamma} + 2 \varepsilon_{\beta\gamma\delta} (E^{\alpha\beta} D^{\delta}B_{\alpha}{}^{\gamma} + 
B^{\alpha\beta} D^{\delta}E_{\alpha}{}^{\gamma})\;,\nonumber\\
&&\label{eq:liedAscalar}\\
&&\mathcal{B}(B)\equiv\pounds_{\vec{\boldsymbol n}} B = -4 (3 B^{\alpha\beta} E_{\alpha}{}^{\gamma} K_{\beta\gamma} -  B K^{\gamma}{}_{\gamma}) + 
2 \varepsilon_{\beta\gamma\delta} (B^{\alpha\beta} D^{\delta}B_{\alpha}{}^{\gamma} -  E^{\alpha\beta} D^{\delta}E_{\alpha}{}^{\gamma})\;,
\label{eq:liedBscalar}
\end{eqnarray}
which we keep for later use. Note that these expressions do not depend on the acceleration $A^\mu$ of the vector field $n^\mu$.

\medskip
The orthogonal splitting of the scalars $\alpha$, $\beta$ and $T$ is found using their definitions and (\ref{eq:splitting-A})-
(\ref{eq:splitting-E}). It is evident that these expressions are functions 
of only $E_{\mu\nu}$, $B_{\mu\nu}$ and their explicit form is not needed in this article.  
Therefore we conclude that $\alpha$, $\beta$ and $T$ can be rendered exclusively in terms of 
the Weyl tensor electric and magnetic parts.

\medskip
The orthogonal splitting of the tensor $Q_{\mu\nu\lambda\rho}$ is written in terms of its electric and magnetic parts just as happens with any other 
Weyl candidate. These are given in terms of the Weyl tensor electric and magnetic parts by
\begin{eqnarray}
&& \mathcal{E}(Q)_{\mu\nu}= \beta E_{\mu\nu} + \alpha B_{\mu\nu}\;, \label{eq:ot-Q0}\\
&& \mathcal{B}(Q)_{\mu\nu}= \beta B_{\mu\nu} - \alpha E_{\mu\nu}.
\label{eq:ot-Q}
\end{eqnarray}
We have now gathered enough information to compute the orthogonal splitting of the different scalar and tensorial quantities
used in Theorem \ref{theorem:kerr-characterization}. The results of these computations are presented in the following series of propositions.
The proofs are straightforward but tedious computations involving the results reviewed in section \ref{section:osplitting} and 
the results just presented about the splitting of the Weyl tensor.

\begin{proposition}
Condition (\ref{eq:typeDcondition}) of Theorem \ref{theorem:kerr-characterization} is equivalent to the following two conditions expressed in terms of 
$E_{\mu\nu}$ and $B_{\mu\nu}$
\begin{eqnarray}
&&\mathfrak{a}_{\mu\nu}\equiv -B{}_{\mu}{}^{\lambda} B{}_{\nu}{}_{\lambda} + E{}_{\mu}{}^{\lambda} E{}_{\nu}{}_{\lambda} 
-\frac{1}{3} A h{}_{\mu}{}_{\nu} -E{}_{\mu}{}_{\nu} \alpha -B{}_{\mu}{}_{\nu} \beta =0\;, \\ 
&&\mathfrak{b}_{\mu\nu}\equiv B{}_{\mu}{}^{\lambda} E{}_{\nu}{}_{\lambda} +  B{}_{\nu}{}^{\lambda} E{}_{\mu}{}_{\lambda}
-\frac{1}{3} B h{}_{\mu}{}_{\nu} -B{}_{\mu}{}_{\nu} \alpha+E{}_{\mu}{}_{\nu}\beta =0.
\end{eqnarray}
 \label{proposition:typeDsplpitting}
\end{proposition}
\proof

Define the tensor
\begin{equation}
 \mathcal{R}_{\mu\nu\lambda\rho}\equiv
\frac{1}{2}\updn{C}{\sigma\tau}{\mu\nu}C_{\sigma\tau\lambda\rho} + \alpha C_{\mu\nu\lambda\rho}+
\beta C^*_{\mu\nu\lambda\rho}- \frac{1}{3}\left( A G_{\mu\nu\lambda\rho}-
B\eta_{\mu\nu\lambda\rho}\right).
\end{equation}
This tensor has the same algebraic monoterm symmetries as the Riemann tensor and therefore
its orthogonal splitting is similar to it. To carry out the proof we 
compute the orthogonal splitting of $\mathcal{R}_{\mu\nu\lambda\rho}$ using its definition
in terms of the Weyl tensor and then equal the resulting expression to (\ref{eq:riemann-split}). 
In this way one obtains the explicit expressions for the tensors 
$X(\mathcal{R})_{\mu\nu}$, $Y(\mathcal{R})_{\mu\nu}$, $Z(\mathcal{R})_{\mu\nu}$ in terms of $E_{\mu\nu}$, $B_{\mu\nu}$. 
These expressions turn out to be
\begin{equation}
 Y(\mathcal{R})_{\mu\nu}=-\mathfrak{a}_{\mu\nu}\;,\quad
 Z(\mathcal{R})_{\mu\nu}=\mathfrak{b}_{\mu\nu}\;,\quad
 X(\mathcal{R})_{\mu\nu}=\mathfrak{a}_{\mu\nu}.
\label{eq:xyz}
 \end{equation}
Since $\mathcal{R}_{\mu\nu\lambda\rho}$ is zero if and only if all its spatial parts vanish then the proposition follows.\qed
\begin{remark}\em
From eq. (\ref{eq:xyz}) we deduce that the only independent spatial parts of $\mathcal{R}_{\mu\nu\lambda\rho}$ are just $\mathfrak{a}_{\mu\nu}$
and $\mathfrak{b}_{\mu\nu}$ which are symmetric and traceless. Therefore the tensor $\mathcal{R}_{\mu\nu\lambda\rho}$ has the same number 
of independent components as the Weyl tensor (10 components) and in fact its orthogonal splitting is rendered as shown in 
(\ref{weyl-1generalized})-(\ref{weyl-2generalized}). From this we can conclude that $\mathcal{R}_{\mu\nu\lambda\rho}$ is indeed a Weyl candidate
something which is not evident from its tensor definition.
\end{remark}

\begin{proposition}
The orthogonal splitting of the quantities $R_\mu$ and $\Theta_\mu$ is
\begin{equation}
\Theta {}_{\mu}=n{}_{\mu} \Theta ^{\parallel } + \Theta^{\bot}{}_{\mu}\;,\quad
R{}_{\mu}=n{}_{\mu} R^{\parallel } + R^{\bot }{}_{\mu}\;,
\label{eq:otthetar}
\end{equation}
where
\begin{eqnarray}
&&\Theta^{\parallel} = \frac{1}{A^2 + B^2}\bigg(\bigl(2 B (E^{\beta\alpha} E^{\gamma}{}_{\alpha} - 
B^{\beta\alpha} B^{\gamma}{}_{\alpha}) + 4 A B^{\beta\alpha} E^{\gamma}{}_{\alpha}\bigr) K_{\beta\gamma} -\nonumber\\
&&\frac{2}{3} \varepsilon_{\beta\gamma\delta} \bigl(E^{\alpha\beta} (B D^{\delta}B_{\alpha}{}^{\gamma} 
+ A D^{\delta}E_{\alpha}{}^{\gamma}) + B^{\alpha\beta} (- A D^{\delta}B_{\alpha}{}^{\gamma} + 
B D^{\delta}E_{\alpha}{}^{\gamma})\bigr)\bigg)\;,\\
&&R^{\parallel} = - \frac{4}{3} K^{\gamma}{}_{\gamma}
+\frac{1}{A^2 + B^2}\bigg(\bigl(2 A (E^{\beta\alpha} E^{\gamma}{}_{\alpha} - B^{\beta\alpha} B^{\gamma}{}_{\alpha}) 
+ 4 B B^{\beta\alpha} E^{\gamma}{}_{\alpha}\bigr) K_{\beta\gamma}-\nonumber\\ 
&&\frac{2}{3} \varepsilon_{\beta\gamma\delta} \bigl(B^{\alpha\beta} (B D^{\delta}B_{\alpha}{}^{\gamma} 
+ A D^{\delta}E_{\alpha}{}^{\gamma}) + E^{\alpha\beta} (A D^{\delta}B_{\alpha}{}^{\gamma} -B D^{\delta}E_{\alpha}{}^{\gamma})\bigr)\bigg)\;,\\
&&R^{\bot }{}_{\mu}\equiv\frac{1}{6}D_{\mu}(\log(A^2+B^2))\;,\\
&&\Theta^{\bot }{}_{\mu}\equiv\frac{1}{3}\left(A^{2}+B^{2}\right)^{-1} (B D{}_{\mu}A- A D{}_{\mu}B).
\label{eq:otRTheta}
\end{eqnarray}
\label{prop:otRTheta}
\end{proposition}
\proof To prove this we need to find the orthogonal splitting of $\nabla_\mu A$ and $\nabla_\mu B$
which is given by
\begin{equation}
 \nabla_\mu A =D_\mu A - n_\mu \pounds_{\vec{\boldsymbol n}}A\;,\quad 
 \nabla_\mu B =D_\mu B - n_\mu \pounds_{\vec{\boldsymbol n}}B\;,
\label{eq:ot-nablaAB}
 \end{equation}
From these relations the values for $R^{\bot }{}_{\mu}$ and $\Theta^{\bot}{}_{\mu}$ follow straightforwardly while for 
$R^{\parallel }$ and $\Theta^{\parallel}$ we get
\begin{equation}
 R^{\parallel }\equiv-\frac{1}{6}\pounds_{\vec{\boldsymbol n}}(\log(A^2+B^2))\;,\quad
\Theta^{\parallel }\equiv\frac{1}{3}\left(A^{2} +B^{2}\right)^{-1} (A\pounds_{\vec{\boldsymbol n}}B-B\pounds_{\vec{\boldsymbol n}} A).
 \end{equation}
The final result follows by using on these expressions (\ref{eq:liedAscalar})-(\ref{eq:liedBscalar}).\qed

Using the result of this proposition we can compute the orthogonal splittings of the scalars $K$ and $\sigma$
\begin{eqnarray}
&& K = \frac{2 R^{\parallel} \Theta^{\parallel} - 2 R^{\bot}{}^{\mu}\Theta^{\bot}_{\mu}}{R^{\parallel}{}^2 -\Theta^{\parallel}{}^2 -
R^{\bot}_{\mu} R^{\bot}{}^{\mu} + \Theta^{\bot}_{\mu} \Theta^{\bot}{}^{\mu}}\;,\quad \label{eq:k-spatial}\\
&& \sigma = \tfrac{1}{4} (-2 + K^2 - 3 K T)^2\times\nonumber \\
&&\left(\frac{R^{\parallel}{}^2 - \Theta^{\parallel}{}^2 -  R^{\bot}_{\mu} R^{\bot}{}^{\mu} + \Theta^{\bot}_{\mu} \Theta^{\bot}{}^{\mu}}
{K^2(1 + K T)^2- (K^2 - 3 K T - 2)^2}- \frac{8\alpha}{(K^2 - 3 K T - 2)^2 - 3 K^2(1 + K T)^2}\right).
\label{eq:sigma-spatial}
\end{eqnarray}
With these relations we can get the orthogonal splitting of $\lambda$ by using (\ref{eq:define-lambda}).

\begin{proposition}
Define the tensors
$$
A_{\nu\rho}\equiv Q_{\mu\nu\lambda\rho} \nabla^\mu A \nabla^\lambda A\;,\quad
B_{\nu\rho}\equiv Q_{\mu\nu\lambda\rho} \nabla^\mu B \nabla^\lambda B.
$$ 
Then we have
\begin{eqnarray}
&&\mathfrak{r}(A)\equiv A_{\nu\mu}n^\nu n^\mu=\mathcal{E}(Q){}_{\mu}{}_{\nu} D{}^{\mu}A D{}^{\nu}A\;,\\
&&\mathfrak{j}(A)_\mu\equiv A_{\nu\rho}n^\nu h^{\rho}_{\phantom{\rho}\mu}=-\varepsilon {}_{\mu}{}_{\kappa}{}_{\pi} \mathcal{B}(Q){}_{\lambda}{}^{\pi}
D{}^{\lambda}A D{}^{\kappa}A -\mathcal{E}(Q){}_{\mu}{}_{\lambda} \mathcal{A}(A)D{}^{\lambda}A,\nonumber\\
&&\mathfrak{t}(A)_{\mu\nu}\equiv A_{\kappa\pi}h^{\kappa}_{\phantom{\kappa}\mu}h^{\pi}_{\phantom{\pi}\nu}= \nonumber\\
&& D{}^{\kappa}A (-2 D{}_{(\nu}A\mathcal{E}(Q){}_{\mu)}{}_{\kappa} + h{}_{\mu}{}_{\nu}\mathcal{E}(Q){}_{\kappa}{}_{\pi} D{}^{\pi}A +  
2\mathcal{A}(A)\mathcal{B}(Q){}_{(\mu}{}^{\pi}\varepsilon{}_{\nu)}{}_{\kappa}{}_{\pi})+\nonumber\\ 
&&+\mathcal{E}(Q){}_{\mu}{}_{\nu} (D{}_{\kappa}A D{}^{\kappa}A + \left(\mathcal{A}(A)\right)^{2}). 
\label{eq:Tdecomposition}
\end{eqnarray}
Similar relations and definitions exist for $T(B)_{\nu\rho}$ (they are just obtained by replacing $A$ by $B$ in the previous relations).
\label{prop:Tdecomposition} 
\end{proposition}
\proof To prove this one computes the orthogonal splitting of the tensor $Q_{\mu\nu\lambda\rho}$ in the way explained when we obtained (\ref{eq:ot-Q})
and uses (\ref{eq:liedAscalar})-(\ref{eq:liedBscalar}).\qed

In view of the previous proposition we have the equivalence
\begin{equation}
Q_{\mu\nu\lambda\rho} \nabla^\mu A \nabla^\lambda A + Q_{\mu\nu\lambda\rho} \nabla^\mu B \nabla^\lambda B=0\Longleftrightarrow
\mathfrak{r}(A)=-\mathfrak{r}(B)\;,\quad
\mathfrak{j}_\mu(A)=-\mathfrak{j}_\mu(B)\;,\quad \mathfrak{t}(A)_{\mu\nu}=-\mathfrak{t}(B)_{\mu\nu}.
\end{equation}

\begin{proposition}
The orthogonal splitting of $\Xi_{\mu\nu}$ is given by
\begin{eqnarray}
&&\mathfrak{r}(\Xi)\equiv\frac{1}{(\lambda^{2}-1)\sqrt{A^{2}+B^{2}}}\times\nonumber\\
&&\big(X(\Pi )_ {\mu\nu}(-\Theta^{\bot}{}^{\mu} \Theta^{\bot}{}^{\nu} + R^{\bot}{}^{\mu}R^{\bot}{}^{\nu}) 
- 2 \mathcal{E}(Q)_{\mu\nu}R^{\bot}{}^{\mu}\Theta^{\bot}{}^{\nu})\big)
\;, \nonumber\\
&&\\
&&\mathfrak{j}(\Xi)_\mu\equiv\frac{-1}{(\lambda^{2}-1)\sqrt{A^{2}+B^{2}}}\times\nonumber\\
&&\bigg(
\Theta^{\bot}{}^{\nu} (\mathcal{E}(Q)_{\mu\nu} R^{\parallel} + X(\Pi )_ {\mu\nu}\Theta^{\parallel} -  
\varepsilon_{\mu\rho\alpha} Z(\Pi )_{\nu}{}^{\alpha} \Theta^{\bot}{}^{\rho}) - \nonumber\\
&&R^{\bot}{}^{\nu} 
\bigl(R^{\parallel} X(\Pi )_{\mu\nu} - \mathcal{E}(Q)_ {\mu\nu} \Theta^{\parallel} + \varepsilon_{\mu\nu\alpha} \
\mathcal{B}(Q)_{\rho}{}^{\alpha} \Theta^{\bot}{}^{\rho} + \varepsilon_{\mu\rho\alpha} 
(- R^{\bot}{}^{\rho} Z(\Pi )_{\nu}{}^{\alpha} + \mathcal{B}(Q)_{\nu}{}^{\alpha} \Theta^{\bot}{}^{\rho})\bigr)
\bigg)\;,\nonumber\\
&&\\
&&\mathfrak{t}(\Xi)_{\mu\nu}\equiv\frac{1}{(\lambda^{2}-1)\sqrt{A^{2}+B^{2}}}\times\nonumber\\
&&\bigg( X(\Pi )^{\rho}{}_{\rho} (R^{\bot}_{\mu} R^{\bot}_{\nu} -  \Theta^{\bot}_{\mu}\Theta^{\bot}_{\nu}) 
- 2 \mathcal{E}(Q)_{\mu\nu} (R^{\parallel} \Theta^{\parallel} + R^{\bot}{}^{\rho} \Theta^{\bot}_{\rho})-\nonumber\\
&&2\varepsilon_{\rho p(\nu}\big(\mathcal{B}(Q)_{\mu)}{}^{\rho} (R^{\bot}{}^{p}\Theta^{\parallel} + R^{\parallel} \Theta^{\bot}{}^{p})+
\mathcal{E}(Q)_{\mu)}{}^{\rho} (- R^{\parallel} R^{\bot}{}^{p} + \Theta^{\parallel} \Theta^{\bot}{}^{p})\big)+\nonumber\\
&&2\mathcal{E}(Q)_{\rho(\nu} (R^{\bot}{}^{\rho} \Theta^{\bot}_{\mu)} + R^{\bot}_ {\mu)}\Theta^{\bot}{}^{\rho})+
2 X(\Pi)_{\rho(\nu}(- R^{\bot}_{\mu)} R^{\bot}{}^{\rho} + \Theta^{\bot}_{\mu)}\Theta^{\bot}{}^{\rho})+\nonumber\\
&& X(\Pi)_{\mu\nu} (R^{\parallel}{}^2 + R^{\bot}_{\rho} R^{\bot}{}^{\rho} - \Theta^{\parallel}{}^2 - \Theta^{\bot}_{\rho} \Theta^{\bot}{}^{\rho})+\nonumber\\
&& h_{\mu\nu} \bigl(X(\Pi )^{\alpha}{}_{\alpha} (- R^{\bot}_{\rho} R^{\bot}{}^{\rho} + \Theta^{\bot}_{\rho} \Theta^{\bot}{}^{\rho}) 
- 2 \mathcal{E}(Q)_{\rho\alpha}R^{\bot}{}^{\rho} \Theta^{\bot}{}^{\alpha} + X(\Pi )_{\rho\alpha} (R^{\bot}{}^{\rho} R^{\bot}{}^{\alpha} -  
\Theta^{\bot}{}^{\rho} \Theta^{\bot}{}^{\alpha})\bigr)\bigg)\;,\nonumber\\
\end{eqnarray}
where
\begin{equation}
X(\Pi)_{\mu\nu}\equiv\mathcal{B}(Q)_{\mu\nu} - h_{\mu\nu} (\alpha^2 + \beta^2).
\label{eq:define-XPi}
\end{equation}
\label{prop:ot-XI}
\end{proposition}

\proof To prove this result we need to compute the orthogonal splitting of the tensors $\Pi_{\mu\nu\rho\lambda}$, $Q_{\mu\nu\rho\lambda}$,
$\Theta_\mu$ and $R_\mu$ (see eq. (\ref{eq:xidefinition})). 
The orthogonal splitting of $Q_{\mu\nu\rho\lambda}$ was worked out in (\ref{eq:ot-Q0})-(\ref{eq:ot-Q}) while that of
$\Theta_\mu$ and $R_\mu$ can be found in eq. (\ref{eq:otthetar}) in Proposition \ref{prop:otRTheta} . The tensor $\Pi_{\mu\nu\rho\lambda}$ has
the same algebraic symmetries as the Riemann tensor. Therefore we use again (\ref{eq:riemann-split})
to compute its orthogonal splitting. In this particular case one has that the corresponding spatial parts 
of $\Pi_{\mu\nu\rho\lambda}$ are
\begin{equation}
 X(\Pi)_{\mu\nu} = \mathcal{B}(Q)_{\mu\nu} - h_{\mu\nu} (\alpha^2 + \beta^2)\;,\quad
 Z(\Pi)_{\mu\nu} = - B_{\mu\nu} \alpha -  E_{\mu\nu} \beta =  - \mathcal{E}(Q)_{\mu\nu}\;,\quad
 Y(\Pi)_{\mu\nu} = -X_{\mu\nu}.
\end{equation}
Hence replacing this into eq. (\ref{eq:riemann-split}) the orthogonal splitting of $\Pi_{\mu\nu\rho\lambda}$
follows. 
Inserting this splitting and those of $Q_{\mu\nu\rho\lambda}$, 
$\Theta_\mu$, $R_\mu$ into (\ref{eq:xidefinition}) the result 
follows after long algebra.\qed

\section{Construction of a positive scalar characterizing the Kerr solution}
\label{sec:positive-scalar}
We use the results of the previous sections to define a positive scalar quantity on a vacuum initial data set 
$(\Sigma, h_{ij}, K_{ij})$ which vanishes if and only if the data development is locally isometric to the Kerr solution. 

The quantities obtained from the orthogonal splittings computed in the previous section can be pull-backed to $\Sigma$ if the foliation generated by the integrable unit normal
is related to the data set $(\Sigma, h_{ij}, K_{ij})$ in the way explained in subsection \ref{subsec:cauchy-problem}. 
This means that all the spatial
scalars and tensors which we defined in the previous sections can be indeed regarded as scalars and tensors defined in a 3-dimensional 
Riemannian manifold $\Sigma$ with metric $h_{ij}$. Consistent with the conventions of subsection \ref{subsec:cauchy-problem} 
we keep the same notation and symbols for them and only replace Greek by Latin indices in the tensors. In the case of scalar quantities
the context will indicate whether they are regarded as space-time quantities or quantities defined on $\Sigma$.

\begin{lemma}
If $\vec{\boldsymbol n}$ is the unit vector orthogonal to the leaves of a foliation related to $\Sigma$ in the way described in
subsection \ref{section:osplitting} and $\vec{\boldsymbol \xi}$ is a vector field on $\mathcal{M}$ then
one has the following equivalence 
\begin{eqnarray}
&&\left.\left(\Xi {}_{\mu}{}_{\nu}
-\left(\frac{\alpha}{1 -3 \lambda^{2}}\right)^{\frac{2}{3}} \xi {}_{\mu} \xi {}_{\nu}\right)\right|_{\phi(\Sigma)}=0\Longleftrightarrow\\
&&
\Omega\equiv\big(\mathfrak{t}(\Xi)_{ij}-M Y_{i} Y_{j}\big)\big(\mathfrak{t}(\Xi)^{ij}-M Y^{i} Y^{j}\big)=0\;,\quad
Y^{2} = \frac{\mathfrak{r}(\Xi)}{M}\;,\quad Y{}_{j}= -\frac{\mathfrak{j}(\Xi)_j}{M Y}\;,
\label{eq:lemma1-equations}
\end{eqnarray}
where 
\begin{eqnarray}
&&M\equiv\left(\frac{\alpha}{1 -3 \lambda^{2}}\right)^{\frac{2}{3}}\;,\\
&&Y\equiv-\phi^*(n_\mu \xi^\mu)\;,  \label{eq:kidlapse}\\
&&Y{}_{j}\equiv\phi^*(h_\mu{}^\nu\xi_\nu)\;,\label{eq:kidshift}
\end{eqnarray}
and $\mathfrak{r}(\Xi)$, $\mathfrak{j}(\Xi)_j$, are the pullbacks to $\Sigma$ of the corresponding space-time quantities defined in
Proposition \ref{prop:ot-XI}. Moreover the quantity $\Omega$ is non-negative. 
\label{lemma:omega}
\end{lemma}
\proof The quantity $\Omega$ is a tensor square defined in a Riemannian manifold so it is trivially non-negative.
Define now the tensor
\begin{equation}
\mathcal{A}_{\mu\nu}\equiv\Xi {}_{\mu}{}_{\nu}
-\left(\frac{\alpha}{1 -3 \lambda^{2}}\right)^{\frac{2}{3}} \xi {}_{\mu} \xi {}_{\nu}.
\end{equation}
The orthogonal splitting of $\mathcal{A}_{\mu\nu}$ is easily computed given that we know the orthogonal splitting of each of 
the terms intervening in its definition (cf. Proposition \ref{prop:ot-XI})
\begin{eqnarray}
&&\mathfrak{r}(\mathcal{A})=-\tilde{Y}^2 M+\mathfrak{r}(\Xi)\;,\quad \tilde{Y}\equiv -n_\mu\xi^\mu \\
&&\mathfrak{j}(\mathcal{A})_\mu=-\mathfrak{j}(\Xi)_\mu-M \tilde{Y} \tilde{Y}_\mu\;,\quad \tilde{Y}_\mu\equiv h_\mu{}^\nu\xi_\nu\\
&&\mathfrak{t}(\mathcal{A})_{\mu\nu}=\mathfrak{t}(\Xi)_{\mu\nu}-M \tilde{Y}_\mu \tilde{Y}_\nu. \label{eq:mathcal-split}
\end{eqnarray}
We use now equation (\ref{eq:spatialp}) to conclude that $\mathcal{A}_{\mu\nu}$ vanish on $\phi(\Sigma)$ if and only if 
the corresponding pull-backed covariant tensors $\mathfrak{r}(\mathcal{A})$, $\mathfrak{j}(\mathcal{A})_j$, $\mathfrak{t}(\mathcal{A})_{ij}$
are zero on $\Sigma$. This entails
\begin{equation}
 Y=\phi^*(\tilde{Y})\;,\quad Y_j=\phi^*(\tilde{Y}_j)\;,\quad 
 Y^2 = \frac{\mathfrak{r}(\Xi)}{M}\;,\quad 
 Y_j = \frac{-\mathfrak{j}(\Xi)_j}{M Y}.
\end{equation}
which are just the last two equations in (\ref{eq:lemma1-equations}).
Finally since $\Sigma$ is a Riemannian manifold, the scalar $\Omega$ is zero if and only if 
\begin{equation}
 \mathfrak{t}(\Xi)_{ij}-M Y_i Y_j=0\;,
\end{equation}
which, according to (\ref{eq:mathcal-split}), is equivalent to $\mathfrak{t}(\mathcal{A})_{ij}=0$.

\qed

We are now ready to present our explicit characterization of a Kerr initial data set.
\begin{theorem}
 Let $(\Sigma, h_{ij}, K_{ij})$ be a vacuum initial data set and assume that on $\Sigma$ the data fulfills the properties
 \begin{equation}
\sigma>0\;,
\label{eq:first-cond-set}
 \end{equation}
where $\sigma$ is defined by (\ref{eq:sigma-spatial}),
and
\begin{equation}
 \mathcal{K}\equiv \big(\mathfrak{r}(B)^2+\mathfrak{j}(B)_i\mathfrak{j}(B)^i+\mathfrak{t}(B)_{ij}\mathfrak{t}(B)^{ij}\big)
(A^2+B^2)> 0\;,
\end{equation}
with $A$, $B$ given by (\ref{eq:splitting-A})-(\ref{eq:splitting-B}) and 
$\mathfrak{r}(B)$, $\mathfrak{j}(B)_i$, $\mathfrak{t}(B)_{ij}$ are defined in Proposition \ref{prop:Tdecomposition}.
Furthermore, with the definitions of Proposition \ref{prop:ot-XI},
introduce two covariant vector fields $Y$, $Y_j$ on $\Sigma$ defined by 
\begin{equation}
 Y^{2} = \frac{\mathfrak{r}(\Xi)}{M}\;,\quad Y{}_{j}= -\frac{\mathfrak{j}(\Xi)_j}{M Y}\;,
\label{eq:killing-splitting-xi}
\end{equation}
where by assumption one has on $\Sigma$ 
\begin{equation}
Y\neq 0.
\label{eq:kid-timelike}
 \end{equation}

Under these conditions define the following non-negative scalar quantity on $\Sigma$
 \begin{eqnarray}
  &&\mathcal{L}\equiv \frac{(\mathfrak{r}(A)+\mathfrak{r}(B))^2+
  (\mathfrak{j}(A)_i+\mathfrak{j}(B)_i)(\mathfrak{j}(A)^i+\mathfrak{j}(B)^i)+(\mathfrak{t}(A)_{ij}+\mathfrak{t}(B)_{ij}) 
  (\mathfrak{t}(A)^{ij}+\mathfrak{t}(B)^{ij})}{\sigma^{14}}+\nonumber\\
  &&\frac{\mathfrak{a}_{ij}\mathfrak{a}^{ij}+\mathfrak{b}_{ij}\mathfrak{b}^{ij}}{\sigma^4}+
  \frac{\big((1-3\lambda^2)\beta + \lambda (3-\lambda^2)\alpha\big)^2}{\sigma^2}+\frac{(\mathfrak{B}_{ij}\mathfrak{B}^{ij})^3}{\sigma^4}+
  \frac{(\mathfrak{C}_{ij}\mathfrak{C}^{ij})^3}{\sigma^7}+\frac{\Omega}{\sigma^2}\;,\nonumber\\
  &&\label{eq:bark}
 \end{eqnarray}
where the intervening quantities are defined in the following places of the text
\begin{itemize}
 \item $\mathfrak{r}(A)$, $\mathfrak{j}(A)_i$, $\mathfrak{t}(A)_{ij}$ are defined in Proposition \ref{prop:Tdecomposition}.
 \item $\mathfrak{a}_{ij}$, $\mathfrak{b}_{ij}$ are defined in Proposition \ref{proposition:typeDsplpitting}.
 \item $\lambda$ is defined in terms of $K$ and $T$ in (\ref{eq:define-lambda}), $K$ is defined in (\ref{eq:k-spatial}),
 $T$ is defined in (\ref{eq:define-T}) in terms of $A$, $B$, $D$, $E$ which in turn are defined in 
 (\ref{eq:splitting-A})-(\ref{eq:splitting-E}).
 \item $\mathfrak{B}_{ij}$, $\mathfrak{C}_{ij}$ are defined in (\ref{eq:kid-scalars}) for the $Y$ and $Y^j$ introduced above.
 \item $\Omega$ is defined in Lemma \ref{lemma:omega}.
 \item The fields $E_{ij}$ $B_{ij}$ on $\Sigma$ are defined by (\ref{eq:eb-to-initial-data0})-(\ref{eq:eb-to-initial-data})
\end{itemize}
All these quantities are defined exclusively from the initial data variables $h_{ij}$, $K_{ij}$, and their covariant
derivatives. 
Under the stated assumptions we have that on $\Sigma$ the scalar $\mathcal{L}\geq 0$ and it vanishes if and only if 
the data $(\Sigma, h_{ij}, K_{ij})$ correspond to Kerr initial data.
\label{theorem:kerr-initial-data}
\end{theorem}
\proof If $(\Sigma, h_{ij}, K_{ij})$ is a vacuum initial data set then the Cauchy development $D(\Sigma)$ is a vacuum
solution and thus the Riemann and the Weyl tensors are the same. Also $\mathcal{L}$ is a sum of squares so it is trivially 
non-negative. If $\mathcal{L}$ vanishes then, since $\Sigma$ is a Riemannian manifold, this is equivalent to the following conditions 
\begin{enumerate}
 \item \label{cond-a} $\mathfrak{r}(A)=-\mathfrak{r}(B)$,
 \item \label{cond-a1} $\mathfrak{j}(A)_i=-\mathfrak{j}(B)_i$,
 \item \label{cond-a2} $\mathfrak{t}(A)_{ij}=-\mathfrak{t}(B)_{ij}$,
 \item \label{cond-d1} $\mathfrak{a}_{ij}=0$,
 \item \label{cond-d2} $\mathfrak{b}_{ij}=0$,
 \item \label{cond-f} $\big((1-3\lambda^2)\beta + \lambda (3-\lambda^2)\alpha\big)=0$,
 \item \label{kiditem-1} $\mathfrak{B}_{ij}=0$,
 \item \label{kiditem-2} $\mathfrak{C}_{ij}=0$.
 \item \label{omega-item} $\Omega=0$.
\end{enumerate}
If \ref{kiditem-1} and \ref{kiditem-2} hold then $\ref{kid-1}$ and $\ref{kid-2}$ hold on $\Sigma$ and the Cauchy development
$\mathcal{D}(\Sigma)$ admits a Killing vector $\xi^\mu$ such that 
$\phi^*(n_\mu\xi^\mu)=-Y$, $\phi^*(h_{\mu\nu}\xi^\nu)=Y_j$ (Theorem \ref{thm:kid}). 
In addition 
from \ref{omega-item} and Lemma \ref{lemma:omega} we deduce 
\begin{equation}
\left.\left(\Xi {}_{\mu}{}_{\nu}-\left(\frac{\alpha}{1 -3 \lambda^{2}}\right)^{\frac{2}{3}} 
\xi {}_{\mu} \xi {}_{\nu}\right)\right|_{\phi(\Sigma)}=0.
\label{eq:xiinit}
\end{equation}

Also \ref{cond-a}-\ref{cond-a2}, \ref{cond-d1}-\ref{cond-d2} and \ref{cond-f} imply each (see eq. (\ref{eq:spatialp}))

\begin{eqnarray}
&&\left.\left(Q_{\mu\nu\lambda\rho} \nabla^\mu A \nabla^\lambda A + Q_{\mu\nu\lambda\rho} \nabla^\mu B \nabla^\lambda B\right)\right|_{\phi(\Sigma)}=0,
\label{eq:conda}\\
&& \left.\left(\frac{1}{2}\updn{C}{\sigma\tau}{\mu\nu}C_{\sigma\tau\lambda\rho} + \alpha C_{\mu\nu\lambda\rho}+
\beta C^*_{\mu\nu\lambda\rho}- \frac{1}{3}\left( A G_{\mu\nu\lambda\rho}-
B\eta_{\mu\nu\lambda\rho}\right)\right)\right|_{\phi(\Sigma)}=0, \label{eq:typeDconditionSigma}\\
&& \big((1-3\lambda^2)\beta + \lambda (3-\lambda^2)\alpha\big)|_{\phi(\Sigma)} =0, \label{eq:startsigma}
\end{eqnarray}

The Lie derivative 
with respect to a Killing vector vanishes for any concomitant of the Weyl tensor and thus we have 

\begin{eqnarray}
&&\pounds_{\vec{\boldsymbol\xi}}\left(\Xi {}_{\mu}{}_{\nu}-\left(\frac{\alpha}{1 -3 \lambda^{2}}\right)^{\frac{2}{3}} \xi {}_{\mu} \xi {}_{\nu}\right)=0,
\label{eq:prop1}\\
&&\pounds_{\vec{\boldsymbol\xi}}\left(Q_{\mu\nu\lambda\rho} \nabla^\mu A \nabla^\lambda A + Q_{\mu\nu\lambda\rho} \nabla^\mu B \nabla^\lambda B\right)=0,
\label{eq:prop2}\\
&&\pounds_{\vec{\boldsymbol\xi}}\left(\frac{1}{2}\updn{C}{\sigma\tau}{\mu\nu}C_{\sigma\tau\lambda\rho} + \alpha C_{\mu\nu\lambda\rho}+
\beta C^*_{\mu\nu\lambda\rho}- \frac{1}{3}\left( A G_{\mu\nu\lambda\rho}-
B\eta_{\mu\nu\lambda\rho}\right)\right)=0,
\label{eq:prop3}\\
&&\pounds_{\vec{\boldsymbol\xi}}\big((1-3\lambda^2)\beta + \lambda (3-\lambda^2)\alpha\big)=0.\label{eq:prop4}
\end{eqnarray}

Combining (\ref{eq:xiinit})-(\ref{eq:startsigma}) with (\ref{eq:prop1})-(\ref{eq:prop4}) we deduce that there exists an open set 
$\mathcal{U}_1\subset D(\Sigma)$, $\phi(\Sigma)\subset\mathcal{U}_1$ in which (\ref{eq:typeDcondition})-(\ref{eq:stationarykilling}) are 
fulfilled as long as the vector field $\vec{\boldsymbol \xi}$ is not tangent to $\phi(\Sigma)$ (if it was, then $\phi(\Sigma)$ would be a characteristic
of the system (\ref{eq:prop1})-(\ref{eq:prop4})). The vector field  $\vec{\boldsymbol \xi}$ is tangent to $\phi(\Sigma)$ at a point if and only if 
$Y=0$ at that point but this is not possible if (\ref{eq:kid-timelike}) holds.
In addition, the hypotheses $\mathcal{K}>0$ and $\sigma > 0$ on the initial data entail the existence of another open set
$\mathcal{U}_2\subset D(\Sigma)$, $\Sigma\subset\mathcal{U}_2$ such that on that set
\begin{equation}
(A^2+ B^2)|_{\Sigma} \neq 0\;, \quad \big(Q_{\mu\nu\lambda\rho} \nabla^\mu B \nabla^\lambda B\big)|_{\Sigma}\neq 0\;,\quad
\sigma >0.
\label{eq:ABneq0}
\end{equation}
We thus deduce that all the conditions of Theorem \ref{theorem:kerr-characterization} 
are met in the open set $\mathcal{U}_1\cap \mathcal{U}_2$ and therefore we conclude that the 
space-time is locally isometric to the Kerr solution in $\mathcal{U}_1\cap \mathcal{U}_2$. 
Reciprocally let us assume that the vacuum data set $(\Sigma, h_{ij}, K_{ij})$ is a Kerr initial data set.
The conditions of Theorem \ref{theorem:kerr-characterization} imply (\ref{eq:xiinit})-(\ref{eq:startsigma}), (\ref{eq:ABneq0}) 
and then the computations carried out 
in propositions \ref{proposition:typeDsplpitting}, \ref{prop:otRTheta}, \ref{prop:Tdecomposition}, \ref{prop:ot-XI} and Lemma \ref{lemma:omega}
in combination with (\ref{eq:spatialp}) entail conditions \ref{cond-a}-\ref{cond-f} and \ref{omega-item} above.
They also imply the condition $\sigma>0$ and $\mathcal{K}>0$ and therefore the scalar $\mathcal{L}$ is 
well-defined for any Kerr initial data set. 
Next we apply Lemma \ref{lemma:omega} and the fact that the vector $\vec{\boldsymbol\xi}$ appearing in (\ref{eq:stationarykilling}) 
of Theorem \ref{theorem:kerr-characterization}
is a Killing vector. This means that $Y$ and $Y^j$ defined by (\ref{eq:kidlapse})-(\ref{eq:kidshift})
fulfill (\ref{kid-1})-(\ref{kid-2}) (see Theorem \ref{thm:kid}) and condition (\ref{eq:killing-splitting-xi}). 
Hence we find that conditions \ref{kiditem-1}-\ref{kiditem-2} are also satisfied for the $Y$ and $Y^j$ defined by (\ref{eq:killing-splitting-xi}).
From this we finally conclude that the scalar $\mathcal{L}$ vanishes on $\Sigma$ under the conditions of the theorem, as desired.
\qed

\begin{remark}\em
 The scalar $\mathcal{L}$ is defined in such a way that it is dimensionless. One could pursue alternative definitions 
 keeping the spirit of Theorem \ref{theorem:kerr-initial-data} resulting in a scalar quantity with dimensions.
\end{remark}

\section{Conclusions}
\label{sec:conclusions}
We have introduced a non-negative scalar quantity $\mathcal{L}$ which is defined for a vacuum initial data set fulfilling 
certain algebraic {\em non-degeneracy conditions} which are included in the hypotheses of Theorem \ref{theorem:kerr-initial-data}. 
The scalar $\mathcal{L}$ is defined exclusively in terms of the 
quantities used to construct the vacuum inita data, namely, a Riemannian manifold, its Riemannian metric and a symmetric tensor defined on it 
which satisfies the {\em vacuum constraint equations} (\ref{Hamiltonian})-(\ref{Momentum}). The  scalar $\mathcal{L}$ is a scalar quantity defined 
at each point of the Riemannian manifold as a {\em concomitant} of the initial data quantities (polynomial expression of the quantities and their covariant
derivatives with respect to the Levi-Civita covariant derivative of the Riemannian metric). And important step in its definition are 
eqs. (\ref{eq:weyl-electric-magnetic})-(\ref{eq:eb-to-initial-data}) because many intermediate expressions are rendered in terms of them. 

The main applications of this scalar are twofold. On one hand it can be used in the study of the non-linear stability problem of the Kerr solution 
to measure how far an initial data set is from the Kerr initial data. As already mentioned the formulation of the non-linear stability problem 
requires the introduction of a notion which enables us to decide when a hypersurface of the space-time is {\em close} in some sense to a
hypersurface embeddable in the Kerr space-time and our scalar fulfills this role in a simple way. 
The second main application arises from numerical relativity. If one carries out a dynamical 
simulation of a collapsing isolated system 
and one wishes to test 
whether the asymptotic evolution in the exterior region of the source is close to the Kerr solution one could compute 
numerically the scalar $\mathcal{L}$ at each slice with constant time in the computational domain of
the simulation and study whether it approaches zero or not (and if it does the convergence rate).
Note that the computation of the scalar $\mathcal{L}$ is totally algorithmic as it only involves 
algebraic and differential manipulations. Therefore its numeric computation should be computationally
less intensive than other computations in which it is necessary to solve numerically partial 
differential equations on the time slices.

\section*{Acknowledgments}
We thank an anonymous referee for his/her reading of a previous version of the manuscript and the improvements he/she  
made. We also thank Dr. Leo Stein for pointing out typos. Finally 
we thank ``Centro de Matem\'atica'' of Minho University for support.


\end{document}